\documentclass[twocolumn]{svjour3}          
\smartqed  
\usepackage{graphicx}
\usepackage{cite}
\usepackage{natbib}
\usepackage{booktabs}
\usepackage{wrapfig}
\usepackage{subfigure}
\usepackage{url}
\usepackage{multicol}
\usepackage{balance}
\usepackage{xcolor}
\journalname{Social Network Analysis and Mining}
\begin{document}

\title{Characterizing the 2016 Russian IRA Influence Campaign
}
\subtitle{}


\author{Adam Badawy         \and
        Aseel Addawood  \and
        Kristina Lerman \and
        Emilio Ferrara
}


\institute{Adam Badawy \at
              USC Information Sciences Institute \\
              \email{abadawy@usc.edu}           
           \and
           Aseel Addawood \at
              University of Illinois at Urbana-Champaign \\
              \email{aaddaw2@illinois.edu}
            \and
           Kristina Lerman \at
              USC Information Sciences Institute \\
              \email{lerman@isi.edu}
            \and
           Emilio Ferrara \at
              USC Information Sciences Institute \\
              \email{emiliofe@usc.edu}
}

\date{Received: date / Accepted: date}

\maketitle

\begin{abstract}
Until recently, social media were seen to promote democratic discourse on social and political issues. However, this powerful communication ecosystem has come under scrutiny for allowing hostile actors to exploit online discussions in an attempt to manipulate public opinion.  A case in point is the  ongoing U.S. Congress investigation of Russian interference in the 2016 U.S. election campaign, with Russia accused of, among other things, using trolls (malicious accounts created for the purpose of manipulation) and bots (automated accounts) to spread  propaganda and politically biased information. In this study, we explore the effects of this manipulation campaign, taking a closer look at users who re-shared the posts produced on Twitter by the Russian troll accounts publicly disclosed by U.S. Congress investigation. We collected a dataset of  13 million election-related posts shared on Twitter in the year of 2016 by over a million distinct users. This dataset includes accounts associated with the identified Russian trolls as well as users sharing posts in the same time period on a variety of topics around the 2016 elections. We use label propagation to infer the users' ideology based on the news sources they share. We are able to classify a large number of the users as liberal or conservative with precision and recall above 84\%.  Conservative  users who retweet Russian trolls produced significantly more tweets than liberal ones, about 8 times as many in terms of tweets. Additionally, trolls' position in the retweet network is stable overtime, unlike users who retweet them who form the core of the election-related retweet network by the end of 2016. Using state-of-the-art bot detection techniques, we estimate that about 5\%  and 11\% of liberal and conservative users are bots, respectively. Text analysis on the content shared by trolls reveal that conservative trolls talk about refugees, terrorism, and Islam; while liberal trolls talk more about school shootings and the police. Although an ideologically broad swath of Twitter users were exposed to Russian trolls in  the period leading up to the 2016 U.S. Presidential election, it is mainly conservatives who help amplify their message.


\end{abstract}

\begin{acknowledgements}
The authors gratefully acknowledge support by the Air Force Office of Scientific Research (award \#FA9550-17-1-0327). The views and conclusions contained herein are those of the authors and should not be interpreted as necessarily representing the official policies or endorsements, either expressed or implied, of AFOSR or the U.S. Government.
\end{acknowledgements}

\newpage

\section{Introduction}
Social media have helped foster democratic conversation about social and political issues: from the Arab Spring \cite{gonzalez2011dynamics}, to Occupy Wall Street movements \cite{conover2013digital, conover2013geospatial} and  other civil protests \cite{gonzalez2013broadcasters, varol2014evolution, stella2018bots}, Twitter and other social media platforms appeared to play an instrumental role in involving the public in policy and political conversations by collectively framing the narratives related to particular social issues, and coordinating online and off-line activities. The use of digital media for political discussions during presidential elections is examined by many  studies, including the past four U.S. Presidential elections \cite{adamic2005political, diakopoulos2010characterizing, bekafigo2013tweets, carlisle2013social, digrazia2013more}, and other countries like Australia \cite{gibson2006does, bruns2011use}, and Norway \cite{enli2013personalized}. Findings that focus on the positive effects of social media, such as increasing voter turnout \cite{bond201261}  or exposure to diverse political views \cite{bakshy2015exposure} contribute to the general praise of these platforms as a tool for promoting democracy and civic engagement \cite{shirky2011political, loader2011networking, effing2011social, tufekci2012social, tufekci2014big}. 

However, concerns regarding the possibility of manipulating public opinion and spreading political propaganda or fake news through social media were also raised early on \cite{howard2006new}. These effects are documented by several studies \cite{ratkiewicz2011detecting, conover2011political, el2013social, woolley2016automation, shorey2016automation, bessi2016social, ferrara2017disinformation, fourney2017geographic}. Social media have been proven to be an effective tool to influence individuals' opinions and behaviors \cite{aral2009distinguishing, aral2012identifying,bakshy2011everyone, centola2011experimental,centola2010spread} and some studies even evaluate the current tools to combat misinformation \cite{Pennycook2017b}. 
Computational tools, like troll accounts and social bots, have been designed to perform such type of influence operations at scale, by cloning or emulating the activity of human users while operating at much higher pace (e.g., automatically producing content following a scripted agenda) \cite{hwang2012socialbots, messias2013you, ferrara2016rise, varol2016online, ferrara2018measuring} -- however, it should be noted that bots have been also used, in some instances, for positive interventions \cite{savage2016botivist, monsted2017evidence}.

Early accounts of the adoption of bots to attempt manipulate political communication with misinformation started in 2010, during the U.S. midterm elections, when social bots were employed to support some candidates and smear others; in that instance, bots injected thousands of tweets pointing to Web sites with fake news \cite{ratkiewicz2011truthy}. Similar cases are reported during the 2010 Massachusetts special election \cite{metaxas2012social} -- these campaigns are often referred to as Twitter bombs, or political astroturf \cite{ferrara2016detection, varol2017early}. Unfortunately, oftentimes determining the actors behind these operations was impossible \cite{kollanyi2016bots, ferrara2016rise}. Prior to this work, only a handful of other operations are linked to some specific actors \cite{woolley2016automation}, e.g., the alt-right attempt to smear a presidential candidate before the 2017 French election \cite{ferrara2017disinformation}. This is because governments, organizations, and other entities with sufficient resources, can obtain the technological capabilities necessary to covertly deploy hundreds or thousands of accounts and use them to either support or attack a given political target. Reverse-engineering these strategies has proven a challenging research venue \cite{freitas2015reverse, alarifi2016twitter, subrahmanian2016darpa, davis2016botornot}, but it can ultimately lead to techniques to identify the actors behind these operations.

One difficulty facing such studies is objectively determining what is fake news, as there is a range of untruthfulness from simple exaggeration to outright lies. Beyond factually wrong information, it is difficult to classify information as fake. 

Rather than facing the conundrum of normative judgment and arbitrarily determine what is fake news and what is not, in this study we focus on user intents, specifically the \textit{intent to deceive}, and their effects on the Twitter political conversation prior to the 2016 U.S. Presidential election.

Online accounts that are created and operated with the primary goal of manipulating public opinion (for example, promoting divisiveness or conflict on some social or political issue) are commonly known as \textit{Internet trolls} (trolls, in short)~\cite{buckels2014trolls}.
To label some accounts or sources of information as trolls, a clear \textit{intent to deceive or create conflict} has to be present. 
A malicious intent to harm the political process and cause distrust in the political system was evident in 2,752 now-deactivated Twitter accounts that are later identified as being tied to Russia's ``Internet Research Agency" troll farm, which was also active on Facebook~\cite{dutt2018senator}. 
The U.S. Congress released a list of these accounts as part of the official investigation of Russian efforts to interfere in the 2016 U.S. Presidential election. 

Since their intent is clearly malicious, the Russian Troll accounts and their messages are the subject of our scrutiny: we study their spread on Twitter to understand the extent of the Russian interference effort and its effects on the election-related political discussion.

\subsection{Research Questions}

In this paper, we aim to answer three crucial research questions regarding the effects of the interference operation carried out by Russian trolls:

\begin{itemize}
\item[RQ1] \textit{What is the role of the users' political ideology?.} We investigate whether political ideology affects who engages with Russian trolls, and how that may have helped propagate trolls' content. If that is the case, we will determine if the effect is more pronounced among liberals or conservatives, or evenly spread across the political spectrum.
\item[RQ2] \textit{How central are the trolls in the network of spreading information in the year of 2016 (pre and post-the US presidential elections)?} We offer analyses of the position of trolls and the users who spread their messages in the retweet network progressively in time from the beginning of 2016 to the end of that year.  
\item[RQ3] \textit{What is the role of social bots?} We characterize whether social bots play a role in spreading content produced by Russian trolls and, if that was the case, where on the political spectrum are bots situated.
\item[RQ4] \textit{Do trolls succeed in specific areas of the US?.} We offer an extensive analysis of the geospatial dimension and how it affects the effectiveness of the Russian interference operation; we test whether users located within specific states participate in the consumption and propagation of trolls' content more than others.
\end{itemize}

This paper improves upon our previous work \cite{Badawy2018}  by 1) extending the span of the data to one year before and after the 2016 US elections rather than just two months as in the previous paper; 2) using sophisticated network analysis to understand the influence of malicious users across time. We collect Twitter data for a year leading into the election. We obtained a  dataset of over 13 million tweets generated by over a million distinct users in the year of 2016. We successfully determine the political ideology of most of the users using label propagation on the retweet network  with precision and recall exceeding 84\%. Next, using advanced machine learning techniques developed to discover social bots \cite{ferrara2016rise, subrahmanian2016darpa, davis2016botornot} applied on users who engage with Russian trolls, we find that bots existed among both liberal and conservative users. We perform text analysis on the content Russian trolls disseminated, and find that conservative trolls are concerned with particular causes, such as refugees, terrorism and Islam, while liberal trolls write about issues related to the police and school shootings. Additionally, we offer an extensive geospatial analysis of tweets across the United States, showing that it is mostly proportionate to the states' population size---as expected---however, a few outliers emerge for both liberals and conservatives.

\subsection{Summary of Contributions}
Findings presented in this work can be summarized as:

\begin{itemize}
\item  We propose a novel way of measuring the consumption of manipulated content through the analysis of activities of Russian trolls on Twitter in the year of 2016.
\item Using network-based machine learning methods, we accurately determine the political ideology of most  users in our dataset, with precision and recall above 84\%.
\item We use network analysis to map the position of trolls and spreaders in the retweet network over the year of 2016.
\item State-of-the-art bot detection on users who engage with Russian trolls shows that bots are engaged in both liberal and conservative domains.
\item Text analysis shows that conservative Russian trolls are mostly promoting conservative causes in regards to refugees, terrorism, and Islam as well as talking about Trump, Clinton, and Obama. For liberal trolls, the discussion is focused on school shootings and the police, but Trump and Clinton are in the top words used as well. 
\item We offer a comprehensive geospatial analysis showing that certain states overly-engaged with production and diffusion of Russian trolls' contents.
\end{itemize}

\section{Data Collection}
\subsection{Twitter Dataset}
\subsubsection{Trolls}

To collect Twitter data about the Russian trolls, we use a list of 2,752 Twitter accounts identified as Russian trolls that is compiled and released by the U.S. Congress.\footnote{https://www.recode.net/2017/11/2/16598312/russia-twitter-trump-twitter-deactivated-handle-list} To collect the tweets, we use Crimson Hexagon,\footnote{https://www.crimsonhexagon.com/} a social media analytic platform that provides paid datastream access. This tool allows us to obtain tweets and retweets produced by trolls and subsequently deleted in the year of 2016. Table \ref{Table 3} offers some descriptive statistics of the Russian troll accounts. Out of the accounts appearing on the list, 1,148 accounts  exist in the dataset, and a little over a thousand of them produce more than half a million of original tweets.

\begin{table}[]
\centering
\caption{Descriptive Statistics of Russian trolls.}
\label{Table 3}
\begin{tabular}{@{}ll@{}}
\toprule
  & Value                                    \\ \midrule
\# of Russian trolls               & 2,735  \\
\# of trolls in our data           & 1,148    \\
\# of trolls wrote original tweets & 1,032     \\
\# of original tweets              & 538,166    \\
\bottomrule
\end{tabular}
\end{table}

We also collect users that did not retweet any troll, since it helps us have a better understanding of trolls behaviour online vs. normal users and how it affects the overall discourse on Twitter. We collect non-trolls' tweets using two strategies. First, we collect tweets of such users using a list of hashtags and keywords that relate to the 2016 U.S. Presidential election. This list is crafted to contain a roughly equal number of hashtags and keywords associated with each major Presidential candidate: we select 23 terms, including five terms referring to the Republican Party nominee Donald J. Trump (\#donaldtrump, \#trump2016, \#neverhillary, \#trumppence16, \#trump), four terms for Democratic Party nominee Hillary Clinton (\#hillaryclinton, \#imwithher, \#nevertrump, \#hillary), and several terms related to debates. To make sure our query list was comprehensive, we add a few keywords for the two third-party candidates, including the Libertarian Party nominee Gary Johnson (one term), and Green Party nominee Jill Stein (two terms).Our second strategy is to collect tweets from the same users that do not include the same key terms mentioned above.

\begin{figure}\centering
  \includegraphics[width=\columnwidth]{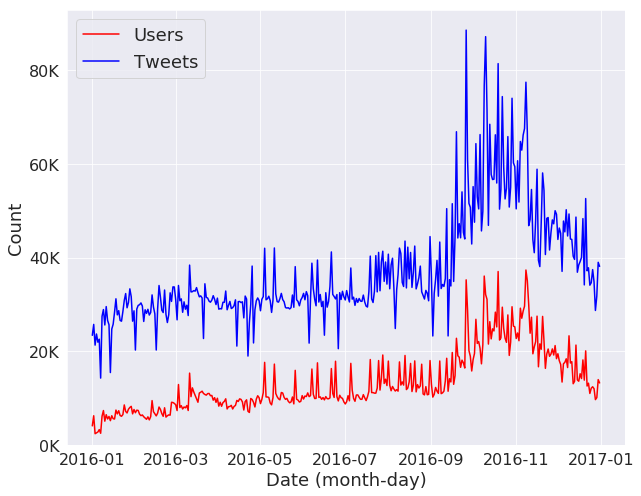}
  \caption{Timeline of the volume of tweets (in blue) generated during our observation period and users the produced these tweets (in red).}
  \label{fig:Timeline}
\end{figure}

Table \ref{Table 1} reports some aggregate statistics of the data. It shows that the number of retweets and tweets with URLs are quite high, more than 3/4 of the dataset. Figure \ref{fig:Timeline} shows the timeline of the tweets' volme and the users who produced these tweets in the 2016 with a spike around the time of the election. 

\begin{table}[tbp]
\centering
\caption{Twitter Data Descriptive Statistics.}
\label{Table 1}
\begin{tabular}{ll}
\toprule
Statistic                         & Count      \\ \midrule
\# of Tweets                     & 13,631,266 \\
\# of Retweets                   & 10,556,421 \\
\# of Distinct Users              & 1,089,974  \\ 
\# of Tweets/Retweets with a URL & 10,621,071 \\ \bottomrule
\end{tabular}
\end{table}

\subsection{Classification of Media Outlets}

We classify users by their ideology based on the political leaning of the media outlets they share. The classification algorithm is described in Section \ref{sec:labelpropagation}. In this section, We describe the methodology of obtaining ground truth labels for the media outlets.
 
We use lists of partisan media outlets compiled by third-party organizations, such as AllSides\footnote{\url{https://www.allsides.com/media-bias/media-bias-ratings}} and Media Bias/Fact Check.\footnote{\url{https://mediabiasfactcheck.com/}} We combine liberal and liberal-center media outlets into one list and conservative and conservative-center into another. The combined list includes 641 liberal and 398 conservative outlets. However, in order to cross reference these media URLs with the URLs in the Twitter dataset, we need to get the long URLs for most of the links in the dataset, since most of them are shortened. As this process is quite time-consuming, we get the top 5,000 URLs by count and then retrieve the long version for those. These top 5,000 URLs accounts for more than 2.1 million or around 1/5 of the URLs in the dataset. 

\begin{figure*}
  \includegraphics[width=0.5\textwidth]{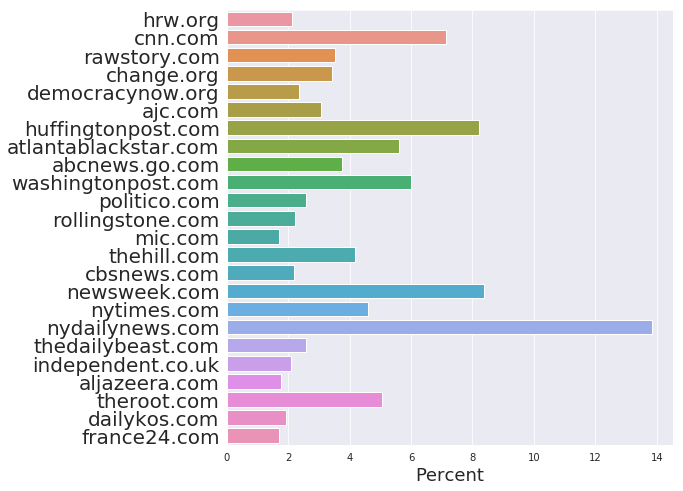}
  \includegraphics[width=0.5\textwidth]{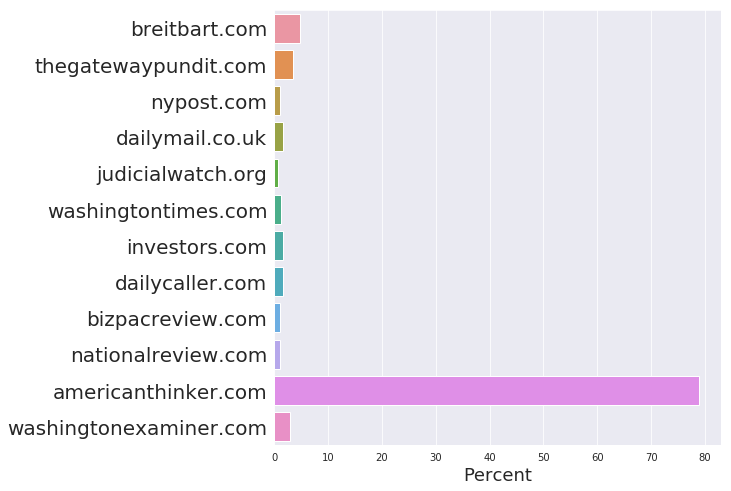}
\caption{Distribution of tweets with links to liberal (right) and conservative (left) media outlets.}
\label{fig:media}   
\end{figure*}

After cross-referencing the 5,000 long URLs with the media URLs, we observe that 7,912  tweets in the dataset contain a URL that points to one of the liberal media outlets and 29,846 tweets with a URL pointing to one of the conservative media outlets. Figure \ref{fig:media}  shows the distributions of tweets with URLs from liberal and conservative outlets. As we can see in the figures, American Thinker dominated the URLs being shared in the conservative sample under study, while in the liberal sample more media outlets were equally represented. Table \ref{Table 2} shows the list of the left and right media outlets/domain names after removing left/right center from the list.

\begin{table}[tbp]
\centering
\caption{Liberal \&  Conservative Domain Names (excluding left-center and right-center)}
\label{Table 2}
\begin{tabular}{@{}ll@{}}
\toprule
Liberal                & Conservative             \\ \midrule
change.org  & dailycaller.com        \\
cnn.com      & nypost.com \\
dailykos.com   & americanthinker.com        \\
democracynow.org          & bizpacreview.com       \\
huffingtonpost.com       & breitbart.com       \\ 
nydailynews.com  & dailymail.co.u        \\
rawstory.com      & judicialwatch.org \\
rollingstone.com   & nationalreview.com        \\
thedailybeast.com          & thegatewaypundit.com      \\
theroot.com       &        \\ \bottomrule
\end{tabular}
\end{table}

We use a polarity rule to label Twitter users as liberal or conservative depending on the number of tweets they produce with links to liberal or conservative sources. In other words, if a user has more tweets with URLs to liberal sources, he/she is labeled as liberal and vice versa. Although the overwhelming majority of users include URLs that are either liberal or conservative, we remove any users that has equal number of tweets from each side. Our final set of labeled users includes 10,074 users.

\section{Methods and Data Analysis}
\subsection{Retweet Network}
 
We construct a retweet network, containing nodes (Twitter users) with a directed link between them if one user retweet a post of another. Table \ref{Table 4} shows the descriptive statistics of the retweet network. It is a sparse network with a giant component that includes 1,357,493 nodes.

\begin{table}[tbp]
\centering
\caption{Descriptive statistics of the Retweet Network.}
\label{Table 4}
\begin{tabular}{@{}ll@{}}
\toprule
Statstic       & Count      \\ \midrule
\# of nodes    & 1,407,190  \\
\# of edges    & 4,874,786 \\
Max in-degree  & 198,262    \\
Max out-degree & 8,458     \\
Density        & 2.46e-06   \\ \bottomrule
\end{tabular}
\end{table}

\subsection{Label Propagation} \label{sec:labelpropagation}

We use label propagation\footnote{We use the algorithm in the Python version of the Igraph library \cite{csardi2006igraph}} to classify Twitter accounts as liberal or conservative, similar to prior work~\cite{Conover2010predicting}. In a network-based label propagation algorithm each node is assigned a label, which is updated iteratively based on the labels of node's network neighbors. In label propagation, a node takes the most frequent label of its neighbors as its own new label. The algorithm proceeds updating labels iteratively and stops when the labels no longer change (see \cite{raghavan2007} for more information). The algorithm takes as parameters (i) weights, in-degree or how many times node $i$ retweets node $j$;  (ii) seeds (the list of labeled nodes). We fix the seeds' labels so they do not change in the process, since this seed list also serves as our ground truth. 

We construct a retweet network where each node corresponds to a Twitter account and a link exists between pairs of nodes when one of them retweets a message posted by the other. 
We use more than 10,000 users mentioned in the media outlets sections as seeds -- those who mainly retweet messages from either the liberal or the conservative media outlets in Figure \ref{fig:media} -- and label them accordingly. We then run label propagation to label the remaining nodes in the retweet network. 

To validate results of the label propagation algorithm, we apply stratified cross (5-fold) validation to the set of more than 10,000 seeds. We train the algorithm on 4/5 of the seed list and see how it performs on the remaining 1/5. The precision and recall scores are around 0.84. Since we combine liberal and liberal-center into one list (same for conservatives), we can see that the algorithm is not only labeling the far liberal or conservative correctly, which is a relatively easier task, but it is performing well on the liberal/conservative center as well.

\begin{table}[]
\centering
\caption{Precision \& Recall scores on seed users}
\label{Table 5}
\begin{tabular}{@{}lll@{}}
\toprule
 & Seed Users       \\ \midrule
Precision  & 0.84    \\
Recall     & 0.84   \\ \bottomrule
\end{tabular}
\end{table}

\subsection{Bot Detection} \label{sec:bots}

Determining whether either a human or a bot controls a social media account has proven a very challenging task \cite{ferrara2016rise, subrahmanian2016darpa, kudugunta2018deep}. We use an openly accessible solution called Botometer (a.k.a. BotOrNot) \cite{davis2016botornot}, consisting of both a public Web site (\url{https://botometer.iuni.iu.edu/})  and a Python API (\url{https://github.com/IUNetSci/botometer-python}), which allows for making this determination with high accuracy. Botometer is a machine-learning framework that extracts and analyses a set of over one thousand features, spanning six sub classes:
\begin{description}
\item [User:] Meta-data features that include the number of friends and followers, the number of tweets produced by the users,profile description and settings.
\item [Friends:] Four types of links are considered here: retweeting, mentioning, being retweeted, and being mentioned. For each group separately, botometer extracts features about language use, local time, popularity, etc. 
\item [Network:] Botometer reconstructs three types of networks: retweet, mention, and hashtag co-occurrence networks. All networks are weighted according to the frequency of interactions or co-occurrences.
\item [Temporal:] Features related to user activity, including average rates of tweet production over various time periods and distributions of time intervals between events.
\item[Content:] Statistics about length and entropy of tweet text and Part-of-Speech (POS) tagging techniques, which identifies different types of natural language components, or POS tags. 
\item[Sentiment:] Features such as: arousal, valence and dominance scores \cite{warriner2013norms}, happiness score \cite{kloumann2012positivity}, polarization and strength \cite{wilson2005recognizing}, and emotion score \cite{agarwal2011sentiment}.
\end{description}

Botometer is trained with thousands of instances of social bots, from simple to sophisticated, yielding an accuracy above 95 percent~\cite{davis2016botornot}. Typically, Botometer returns likelihood scores above 50 percent only for accounts that look suspicious to a scrupulous analysis. We adopted the Python Botometer API to systematically inspect the most active users in our dataset. The Python Botometer API queries the Twitter API to extract 300 recent tweets and publicly available account metadata, and feeds these features to an ensemble of machine learning classifiers, which produce a bot score.
To label accounts as bots, we use the fifty-percent threshold -- which has proven effective in prior studies \cite{davis2016botornot}: an account is considered to be a bot if the overall Botometer score is above 0.5. 

\subsection{Geolocation}

There are two ways to identity the location of tweets produced by users. One way is to collect the  coordinates of the location the tweets were sent from; however, this is only possible if users enable the geolocation option on their Twitter accounts. The second way is to analyze the self-reported location text in users' profiles. The latter includes substantially more noise, since many people write fictitious or imprecise locations -- for example, they may identify the state and the country they reside in, but not the city.

In this paper, we use the location provided by Crimson Hexagon, which uses two methodologies. First, it extracts the geotagged locations, which is only available in a small portion of the Twitter data. For the tweets which are not geotagged, Crimson Hexagon estimates the users' countries, regions, and cities based on ``various pieces of contextual information'', for example, their profile information as well as users' time zones and languages. Using the country and state information provided by Crimson Hexagon, this dataset has over 9 Million tweets with country location. More than 7.5 million of these geolocated tweets come from the US, with the UK, Nigeria, and Russia trailing behind the US with 287k, 277k, and 192k tweets respectively. There are more than 4.7 Million US tweets with State information provided. The top four states are as expected: California, Texas, New York, and Florida with 742k, 580k, 383k, and 360k tweets.

\section{Results}

Let us address the three research questions we sketched earlier:

\begin{itemize}
\item[RQ1] \textit{What is the role of the users' political ideology?}
\item[RQ2] \textit{How central are trolls and the users who retweet them, spreaders?} 
\item[RQ3] \textit{What is the role of social bots?} 
\item[RQ4] \textit{Did trolls especially succeed in specific areas of the US?} 
\end{itemize}

In Section \ref{sub:pol}, we analyze how political ideology affects engagement with content posted by Russian trolls. Section \ref{sub:temp} shows how the position of trolls and spreaders evolve over time in the retweet network. Section \ref{sub:bots} focuses on social bots and how they contribute to the spreading of content produced by Russian trolls. Finally, in Section \ref{sub:geo} we show how users contribute to the consumption and the propagation of trolls' content based on their location.

\subsection{RQ1: Political Ideology} \label{sub:pol}

The label propagation method succeeds in labeling most of the users as either liberal or conservative; however, the algorithm was not able to label some users outside of the giant component. Table  \ref{Table:trollPiDCount} shows the number of trolls by ideology, and we can see that there are almost double the amount of conservative trolls compared to liberal ones in terms of number of trolls overall and the number of trolls who wrote original tweets. Although the number of liberal trolls is twice smaller, they produce more tweets than conservative trolls. 

Table \ref{Table: spreadersStats} shows descriptive statistics of spreaders. The table shows that only few spreaders wrote original tweets and that more than half of the tweets are retweets of trolls. There are fewer conservative spreaders, but they write substantially more tweets than their liberal counterparts (see Table \ref{Table: PId&BotAnalysis}). Besides talking about the candidates, liberals talk about being black, women, and school shootings. Conservatives talk about being American, Obama, terrorism, refugees, and Muslims (see Table \ref{Table: Top20words} for top 20 words used for liberals and conservatives). 


\begin{table}[]
\centering
\caption{Breakdown of the Russian trolls by political ideology, with the ratio of conservative to liberal trolls.}
\label{Table:trollPiDCount}
\begin{tabular}{@{}llll@{}}
\toprule
& Liberal                        & Conservative & Ratio     \\ \midrule
\# of trolls                   & 339          & 688         & 2   \\
\# of trolls w/ original tweets & 306           & 608          & 1.98 \\
\# of original tweets          & 299,464            & 215,617         & 0.7  \\ \bottomrule
\end{tabular}
\end{table}

\begin{table}[h]
\centering
\caption{Descriptive statistics of spreaders, i.e., users who retweeted Russian trolls.}
\label{Table: spreadersStats}
\begin{tabular}{@{}ll@{}}
\toprule
  & Value                                                          \\ \midrule
\# of spreaders                      & 720,558                   \\
\# of times retweeted trolls    & 3,540,717                   \\ 
\# of spreaders with original tweets & 21,338                   \\
\# of original tweets                & 319,565 \\
\# of original tweets and retweets   & 7,357,717  \\ \bottomrule
\end{tabular}
\end{table}


\begin{table}[]
\centering
\caption{All spreaders by political ideology; Bot analysis for 115,396 spreaders (out of a 200k random sample of  spreaders). Ratio: conservative/liberal}
\label{Table: PId&BotAnalysis}
\begin{tabular}{@{}llll@{}}
\toprule
& Liberal              & Conservative & Ratio    \\ \midrule
\# of spreaders      & 446,979        & 273,546 & 0.6 \\
\# of tweets         & 1,715,696       & 5,641,988  & 3.2 \\ \midrule
\# of bots           & 3,528           & 4,896       & 1.4 \\
\# of tweets by bots & 26,233       & 181,604     & 7 \\ \bottomrule
\end{tabular}
\end{table}

\begin{table}[]
\centering
\caption{Top 20 meaningful lemmatized words from the tweets of Russians trolls classified as Conservative and Liberal}
\label{Table: Top20words}
\begin{tabular}{@{}llll@{}}
\toprule
Conservative & count & Liberal & Count \\ \midrule
trump      & 10,362              & police     & 15,498         \\
hillary    & 5,494               & trump      & 13,999         \\
people     & 4,479               & man        & 13,118         \\
clinton    & 4,295               & black      & 9,942          \\
obama      & 2,593               & year       & 8,627          \\
one        & 2,447               & state      & 7,895          \\
american   & 2,388               & woman      & 7,748          \\
woman      & 2,167               & shooting   & 6,564          \\
day        & 2,152               & killed     & 6,407          \\
donald     & 2,148               & people     & 5,913          \\
time       & 2,105               & clinton    & 5,606          \\
refugee    & 2,103               & school     & 5,143          \\
president  & 2,079               & shot       & 5,132          \\
terrorist  & 1,994               & city       & 4,754          \\
country    & 1,980               & win        & 4,543          \\
muslim     & 1,963               & cop        & 4,515          \\
year       & 1,893               & day        & 4,478          \\
need       & 1,856               & fire       & 4,408          \\
think      & 1,835               & officer    & 4,397          \\
al         & 1,823               & death      & 4,305          \\ \bottomrule
\end{tabular}
\end{table}

\subsection{RQ2: Temporal Analysis} \label{sub:temp}

Analyzing the influence of trolls across time is one of the most important questions in terms of the spread of political propaganda. The way we measure the influence of trolls is by measuring where they are located in the retweet network. We choose the retweet network in particular because retweeting is the main vehicle for spreading information on Twitter. In terms of the location of the user, there are multiple ways to measure the location of a user in the network he/she is embedded in. We choose the k-core decomposition technique, because it captures the notion of who is in the core of the network vs. the periphery, while giving an ordinal measure reflecting the number of connections. 

The k-core of a graph is the maximal subgraph in which every node has at least degree k. In a directed network, k is the sum of in-degree + out-degree. The k-core decomposition is a recursive approach that progressively trims the least connected nodes in a network (those with lower degree) in order to identify the most central ones \cite{barbera2015critical}. 

We measure the centrality of trolls across time by dividing the number of trolls by the total number of nodes in every core for every snapshot of the network. Since we want to measure the evolution of the trolls' importance, we construct monthly retweet networks, where every network contains all the previous nodes and edges up to the time in question. The first snapshot of the network contains the nodes and edges observed until one month after the initialization of the data collection, while the last snapshot includes all nodes and edges in the retweet network. 

\begin{figure*}

  \includegraphics[width=0.5\textwidth]{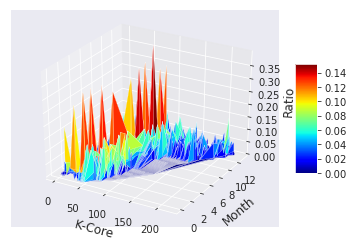}
  \includegraphics[width=0.5\textwidth]{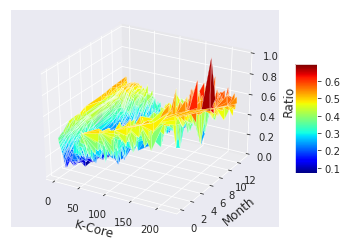}

\caption{Temporal K-Core plots for trolls (left) and spreaders (right); the z-axis show the fraction of  trolls/spreaders over the total per k-core for each month in the retweet network. For each month, the graph contains all nodes and edges from all previous months; in other words, the temporal graphs are additive in terms of nodes and edges.}
\label{fig:3-D Plot}      
\end{figure*}

Figure \ref{fig:3-D Plot} (left side) shows that trolls make up a substantial portion of the lower cores and remain fairly stable across time. We replicate the analysis for spreaders on figure \ref{fig:3-D Plot} (right-side), which shows that spreaders' centrality increases progressively across time and dominates the core of the network toward the end.

\subsection{RQ3: Social Bots} \label{sub:bots}

Since there are many spreaders in this dataset, we take a random sample of  200,000 spreaders and use the approach explained in Section \ref{sec:bots} to obtain bot scores. We use Botometer to obtain bot scores only for the sample and not for all of the spreaders, since it takes considerably a long time to get bot scores for such a number due to the Twitter API call limit. We are able to obtain bot scores for 115,396 spreaders out of the 200,000 spreaders. 

The number of users that have bot scores above 0.5, and can therefore be safely considered bots according to prior work~\cite{varol2016online}, stands at 8,424 accounts. 
Out of the 115,396 spreaders with bot scores, 68,057 are liberal, and 3,528 of them have bot scores above 0.5, about 5\% of the total. As for the conservatives, there are 45,807 spreaders, 4,896 of which have bot scores greater than 0.5, representing around 11\% of the total. As the Results summarized in Table \ref{Table: PId&BotAnalysis} show, although the number of conservative spreaders is lower than liberal ones, there are more bots among conservatives who write considerably more tweets.

\begin{figure}
  \includegraphics[width=\columnwidth]{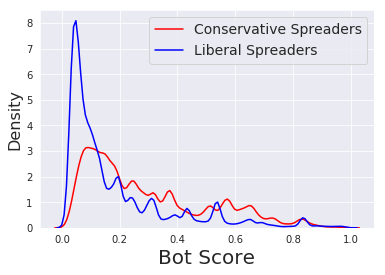}
\caption{Overall Bot Score}
\label{fig:Overall Bot Score}       
\end{figure}

Figure \ref{fig:Overall Bot Score} shows the probability density of bot scores of the liberal and conservative spreaders. Again, putting aside the disproportionate number of liberals to conservatives, the mean value of the bot scores of the conservative spreaders (0.29) is higher than the liberal one (0.18). We performed a two-sided t-test for the null hypothesis that the two distributions have identical mean values, and the p-value is less than 0.0, meaning that we can reject the null.

\begin{figure*}
  \includegraphics[width=0.33\textwidth]{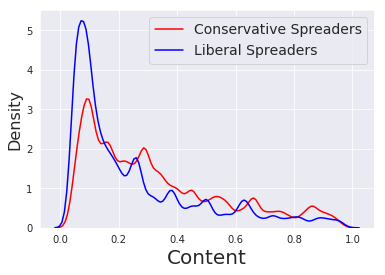}
  \includegraphics[width=0.33\textwidth]{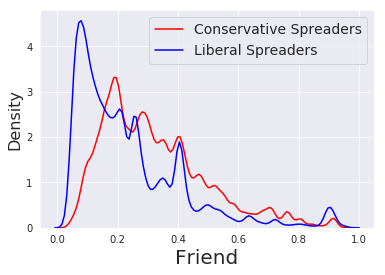}
  \includegraphics[width=0.33\textwidth]{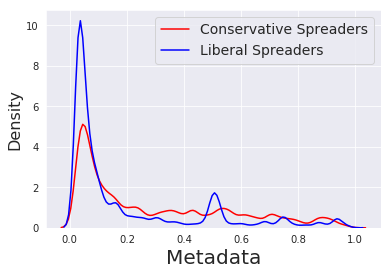}
  \medskip
  \includegraphics[width=0.33\textwidth]{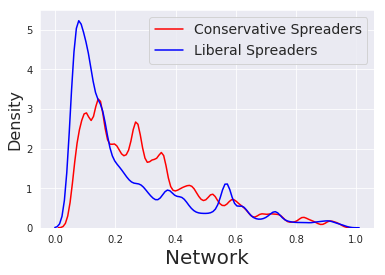}
  \includegraphics[width=0.33\textwidth]{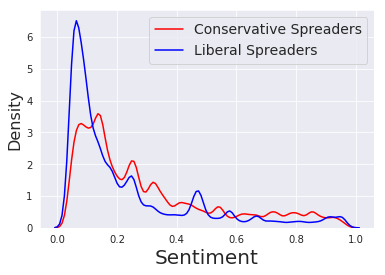}
  \includegraphics[width=0.33\textwidth]{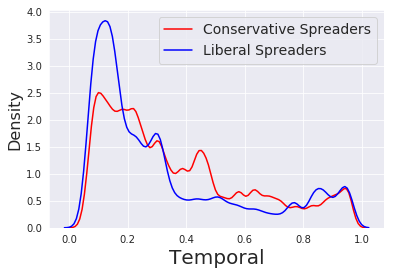}
\caption{Bot Scores by sub-category}
\label{fig:Bot Scores by sub-category}      
\end{figure*}

For the plots in Figure \ref{fig:Bot Scores by sub-category}, it is evident that conservative spreaders have higher means on all of the Botometer subclass scores in comparison to their liberal counterparts. The differences in all plots are statistically significant (p $<$0.001). Besides, looking at the distributions, we can see that the differences in user characteristics (metadata) and ``Friend'' category are substantively different (around 0.10 difference).

\subsection{RQ4: Geospatial Analysis} \label{sub:geo}

\begin{figure*}
  \includegraphics[width=0.5\textwidth,clip=true, trim=0 70 0 80]{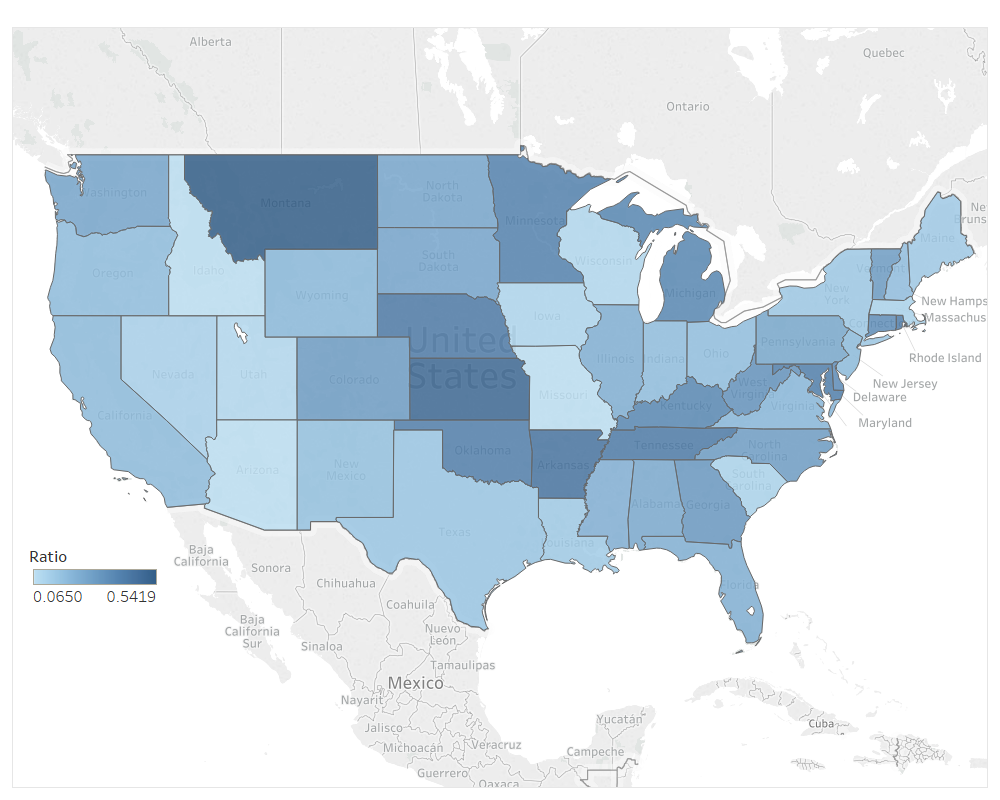}
  \includegraphics[width=0.5\textwidth,clip=true, trim=0 70 0 80]{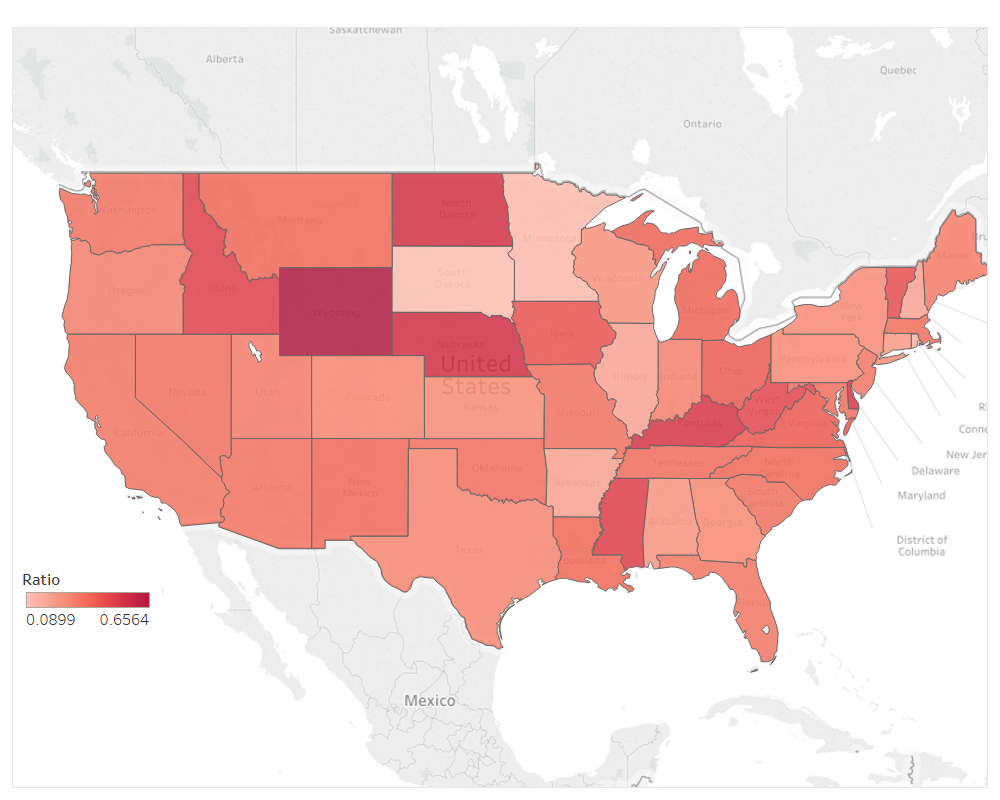}

\caption{Proportion of the number of retweets by conservative users of Russian trolls per each state normalized by the total number of conservative tweets by state(right); equivalent map for liberals on the left.}
\label{fig:maps}       
\end{figure*}

Figure \ref{fig:maps} shows the proportion of the number of retweets by liberal and conservative users (classified according to the label propagation algorithm) of the content produced by Russian trolls per each state normalized per state by the total number of liberal and conservative tweets respectively.  The ratio $\rho$ is computed as $\rho = (T_S / P_S)$, where $T_S$ is the total number of liberal/conservative retweets of liberal/conservative trolls from a given state $S$, and $P_S$ is the total number of tweets per each State.

We notice that few states exhibit very high proportions of retweets per total number of tweets for liberals and conservatives. We test the deviations using a two-tailed t-test on the z-scores of each deviation calculated on the distribution of ratios. For conservatives, the average is 0.34 and the standard deviation is 0.12, while for liberal states, the average is 0.26 and the standard deviation is 0.13. For conservatives, Wyoming leads the ranking ($\rho$=3.65, p-value = 0.001). For liberals, Montana ($\rho$=0.54, p-value = 0.023) and Kansas ($\rho$=0.51, p-value = 0.046) lead the ranking and are the only states with ratios that are statistically significant. It is also interesting to note that a little less than 1/2 of the trolls have Russia as their location while a small number of spreaders and other users have Russia as their location (see Table \ref{Table:Russialocation}).

\begin{table}[]
\centering
\caption{Users that reported Russia as their location}
\label{Table:Russialocation}
\begin{tabular}{@{}ll@{}}
\toprule
& Count                            \\ \midrule
\# of trolls from Russia    & 438  \\
\# of spreaders from Russia & 1,980 \\
\# of overall from Russia   & 3,017 \\ \bottomrule
\end{tabular}
\end{table}

\section{Conclusions}
The dissemination of information and the mechanisms for democratic discussion have radically changed since the advent of digital media, especially social media. Platforms like Twitter are praised for their contribution to democratization of public discourse on civic and political issues. However, many studies highlight the perils associated with the abuse of these platforms. The spread of deceptive, false and misleading information aimed at manipulating public opinion are among those risks.

In this work, we investigated the role and effects of trolls, using the content produced by Russian trolls on Twitter as a proxy for propaganda. We collected tweets posted during the year of 2016 by Russian trolls, spreaders, and other users who tweet in this period.   We showed that propaganda (produced by Russian trolls) is shared more widely by conservatives than liberals on Twitter. Although the number of liberal spreaders is close to 2:1 in comparison to conservative ones, the latter write about 3.2 times as many tweets as the liberal spreaders.  Using  state-of-the-art bot detection method, we estimated that about 5\%  and 11\% of the liberal and conservative users are bots. Conservative bot spreaders produce about 7 times as many tweets as liberal ones. 

The spread of propaganda by malicious actors can have severe negative consequences. It can enhance malicious information and polarize political conversation, causing confusion and social instability. Scientists are currently investigating the consequences of such phenomena \cite{woolley2016automation,shorey2016automation}. We plan to explore in detail the issue of how malicious information spread via exposure and the role of peer effect. 
Concluding, it is important to stress that, although our analysis unveiled the current state of the political debate and agenda pushed by the Russian trolls who spread malicious information, it is impossible to account for all the malicious efforts aimed at manipulation during the last presidential election. State- and non-state actors, local and foreign governments, political parties, private organizations, and even individuals with adequate resources~\cite{kollanyi2016bots}, could obtain operational capabilities and technical tools to construct propaganda campaigns and deploy armies of social bots to affect the directions of online conversations. Future efforts will be required by the social and computational science communities to study this issue in depth and develop more sophisticated detection techniques capable of unmasking and fighting these malicious efforts.




\balance
\bibliographystyle{natbib,abbrv}
\bibliography{citations}

\begin{thebibliography}{}

\bibitem[Adamic and Glance(2005)Adamic and Glance]{adamic2005political}
Adamic, L.~A. and Glance, N. (2005).
\newblock The political blogosphere and the 2004 us election: divided they
  blog.
\newblock In {\em 3rd Int Work Link Disc\/}.

\bibitem[Agarwal {\em et~al.}(2011)Agarwal, Xie, Vovsha, Rambow, and
  Passonneau]{agarwal2011sentiment}
Agarwal, A., Xie, B., Vovsha, I., Rambow, O., and Passonneau, R. (2011).
\newblock Sentiment analysis of twitter data.
\newblock In {\em Proceedings of the workshop on languages in social media\/},
  pages 30--38. Association for Computational Linguistics.

\bibitem[Alarifi {\em et~al.}(2016)Alarifi, Alsaleh, and
  Al-Salman]{alarifi2016twitter}
Alarifi, A., Alsaleh, M., and Al-Salman, A. (2016).
\newblock Twitter turing test: Identifying social machines.
\newblock {\em Information Sciences\/}, {\bf 372}.

\bibitem[Aral and Walker(2012)Aral and Walker]{aral2012identifying}
Aral, S. and Walker, D. (2012).
\newblock Identifying influential and susceptible members of social networks.
\newblock {\em Science\/}, {\bf 337}(6092), 337--341.

\bibitem[Aral {\em et~al.}(2009)Aral, Muchnik, and
  Sundararajan]{aral2009distinguishing}
Aral, S., Muchnik, L., and Sundararajan, A. (2009).
\newblock Distinguishing influence-based contagion from homophily-driven
  diffusion in dynamic networks.
\newblock {\em Proceedings of the National Academy of Sciences\/}, {\bf
  106}(51).

\bibitem[Badawy {\em et~al.}(2018)Badawy, Ferrara, and Lerman]{Badawy2018}
Badawy, A., Ferrara, E., and Lerman, K. (2018).
\newblock Analyzing the digital traces of political manipulation: The 2016
  russian interference twitter campaign.
\newblock In {\em ASONAM\/}.

\bibitem[Bakshy {\em et~al.}(2011)Bakshy, Hofman, Mason, and
  Watts]{bakshy2011everyone}
Bakshy, E., Hofman, J., Mason, W., and Watts, D. (2011).
\newblock Everyone's an influencer: quantifying influence on twitter.
\newblock In {\em 4th WSDM\/}.

\bibitem[Bakshy {\em et~al.}(2015)Bakshy, Messing, and
  Adamic]{bakshy2015exposure}
Bakshy, E., Messing, S., and Adamic, L.~A. (2015).
\newblock Exposure to ideologically diverse news and opinion on facebook.
\newblock {\em Science\/}, {\bf 348}(6239).

\bibitem[Barber{\'a} {\em et~al.}(2015)Barber{\'a}, Wang, Bonneau, Jost,
  Nagler, Tucker, and Gonz{\'a}lez-Bail{\'o}n]{barbera2015critical}
Barber{\'a}, P., Wang, N., Bonneau, R., Jost, J.~T., Nagler, J., Tucker, J.,
  and Gonz{\'a}lez-Bail{\'o}n, S. (2015).
\newblock The critical periphery in the growth of social protests.
\newblock {\em PloS one\/}, {\bf 10}(11), e0143611.

\bibitem[Bekafigo and McBride(2013)Bekafigo and McBride]{bekafigo2013tweets}
Bekafigo, M.~A. and McBride, A. (2013).
\newblock Who tweets about politics? political participation of twitter users
  during the 2011 gubernatorial elections.
\newblock {\em Social Science Computer Review\/}, {\bf 31}(5), 625--643.

\bibitem[Bessi and Ferrara(2016)Bessi and Ferrara]{bessi2016social}
Bessi, A. and Ferrara, E. (2016).
\newblock Social bots distort the 2016 us presidential election online
  discussion.
\newblock {\em First Monday\/}, {\bf 21}(11).

\bibitem[Bond {\em et~al.}(2012)Bond, Fariss, Jones, Kramer, Marlow, Settle,
  and Fowler]{bond201261}
Bond, R.~M., Fariss, C.~J., Jones, J.~J., Kramer, A.~D., Marlow, C., Settle,
  J.~E., and Fowler, J.~H. (2012).
\newblock A 61-million-person experiment in social influence and political
  mobilization.
\newblock {\em Nature\/}, {\bf 489}(7415).

\bibitem[Bruns and Burgess(2011)Bruns and Burgess]{bruns2011use}
Bruns, A. and Burgess, J.~E. (2011).
\newblock The use of twitter hashtags in the formation of ad hoc publics.
\newblock In {\em 6th ECPR General Conference\/}.

\bibitem[Buckels {\em et~al.}(2014)Buckels, Trapnell, and
  Paulhus]{buckels2014trolls}
Buckels, E.~E., Trapnell, P.~D., and Paulhus, D.~L. (2014).
\newblock Trolls just want to have fun.
\newblock {\em Personality and individual Differences\/}, {\bf 67}.

\bibitem[Carlisle and Patton(2013)Carlisle and Patton]{carlisle2013social}
Carlisle, J.~E. and Patton, R.~C. (2013).
\newblock Is social media changing how we understand political engagement? an
  analysis of facebook and the 2008 presidential election.
\newblock {\em Political Research Quarterly\/}, {\bf 66}(4).

\bibitem[Centola(2010)Centola]{centola2010spread}
Centola, D. (2010).
\newblock The spread of behavior in an online social network experiment.
\newblock {\em Science\/}, {\bf 329}(5996), 1194--1197.

\bibitem[Centola(2011)Centola]{centola2011experimental}
Centola, D. (2011).
\newblock An experimental study of homophily in the adoption of health
  behavior.
\newblock {\em Science\/}, {\bf 334}(6060), 1269--1272.

\bibitem[Conover {\em et~al.}(2011a)Conover, Ratkiewicz, Francisco,
  Gon{\c{c}}alves, Menczer, and Flammini]{conover2011political}
Conover, M., Ratkiewicz, J., Francisco, M.~R., Gon{\c{c}}alves, B., Menczer,
  F., and Flammini, A. (2011a).
\newblock Political polarization on twitter.
\newblock {\em ICWSM\/}, {\bf 133}, 89--96.

\bibitem[Conover {\em et~al.}(2011b)Conover, Gon\c{c}alves, Ratkiewicz,
  Flammini, and Menczer]{Conover2010predicting}
Conover, M., Gon\c{c}alves, B., Ratkiewicz, J., Flammini, A., and Menczer, F.
  (2011b).
\newblock Predicting the political alignment of twitter users.
\newblock In {\em Proc. 3rd IEEE Conference on Social Computing\/}, pages
  192--199.

\bibitem[Conover {\em et~al.}(2013a)Conover, Ferrara, Menczer, and
  Flammini]{conover2013digital}
Conover, M.~D., Ferrara, E., Menczer, F., and Flammini, A. (2013a).
\newblock The digital evolution of occupy wall street.
\newblock {\em PloS one\/}, {\bf 8}(5), e64679.

\bibitem[Conover {\em et~al.}(2013b)Conover, Davis, Ferrara, McKelvey, Menczer,
  and Flammini]{conover2013geospatial}
Conover, M.~D., Davis, C., Ferrara, E., McKelvey, K., Menczer, F., and
  Flammini, A. (2013b).
\newblock The geospatial characteristics of a social movement communication
  network.
\newblock {\em PloS one\/}, {\bf 8}(3), e55957.

\bibitem[Csardi and Nepusz(2006)Csardi and Nepusz]{csardi2006igraph}
Csardi, G. and Nepusz, T. (2006).
\newblock The igraph software package for complex network research.
\newblock {\em InterJournal, Complex Systems\/}, {\bf 1695}(5), 1--9.

\bibitem[Davis {\em et~al.}(2016)Davis, Varol, Ferrara, Flammini, and
  Menczer]{davis2016botornot}
Davis, C.~A., Varol, O., Ferrara, E., Flammini, A., and Menczer, F. (2016).
\newblock Botornot: A system to evaluate social bots.
\newblock In {\em Proc. 25th International Conference on World Wide Web\/},
  pages 273--274.

\bibitem[Diakopoulos and Shamma(2010)Diakopoulos and
  Shamma]{diakopoulos2010characterizing}
Diakopoulos, N.~A. and Shamma, D.~A. (2010).
\newblock Characterizing debate performance via aggregated twitter sentiment.
\newblock In {\em SIGCHI Conf.}

\bibitem[DiGrazia {\em et~al.}(2013)DiGrazia, McKelvey, Bollen, and
  Rojas]{digrazia2013more}
DiGrazia, J., McKelvey, K., Bollen, J., and Rojas, F. (2013).
\newblock More tweets, more votes: Social media as a quantitative indicator of
  political behavior.
\newblock {\em PloS one\/}, {\bf 8}(11), e79449.

\bibitem[Dutt {\em et~al.}(2018)Dutt, Deb, and Ferrara]{dutt2018senator}
Dutt, R., Deb, A., and Ferrara, E. (2018).
\newblock {`Senator, We Sell Ads': Analysis of the 2016 Russian Facebook Ads
  Campaign}.
\newblock In {\em Third International Conference on Intelligent Information
  Technologies (ICIIT 2018)\/}.

\bibitem[Effing {\em et~al.}(2011)Effing, Van~Hillegersberg, and
  Huibers]{effing2011social}
Effing, R., Van~Hillegersberg, J., and Huibers, T. (2011).
\newblock Social media and political participation: are facebook, twitter and
  youtube democratizing our political systems?
\newblock {\em Electronic participation\/}, pages 25--35.

\bibitem[El-Khalili(2013)El-Khalili]{el2013social}
El-Khalili, S. (2013).
\newblock Social media as a government propaganda tool in post-revolutionary
  egypt.
\newblock {\em First Monday\/}, {\bf 18}(3).

\bibitem[Enli and Skogerb{\o}(2013)Enli and Skogerb{\o}]{enli2013personalized}
Enli, G.~S. and Skogerb{\o}, E. (2013).
\newblock Personalized campaigns in party-centred politics: Twitter and
  facebook as arenas for political communication.
\newblock {\em Information, Communication \& Society\/}, {\bf 16}(5), 757--774.

\bibitem[Ferrara(2017)Ferrara]{ferrara2017disinformation}
Ferrara, E. (2017).
\newblock Disinformation and social bot operations in the run up to the 2017
  french presidential election.
\newblock {\em First Monday\/}, {\bf 22}(8).

\bibitem[Ferrara(2018)Ferrara]{ferrara2018measuring}
Ferrara, E. (2018).
\newblock Measuring social spam and the effect of bots on information diffusion
  in social media.
\newblock In {\em Complex Spreading Phenomena in Social Systems\/}, pages
  229--255. Springer, Cham.

\bibitem[Ferrara {\em et~al.}(2016a)Ferrara, Varol, Menczer, and
  Flammini]{ferrara2016detection}
Ferrara, E., Varol, O., Menczer, F., and Flammini, A. (2016a).
\newblock Detection of promoted social media campaigns.
\newblock In {\em Tenth International AAAI Conference on Web and Social
  Media\/}, pages 563--566.

\bibitem[Ferrara {\em et~al.}(2016b)Ferrara, Varol, Davis, Menczer, and
  Flammini]{ferrara2016rise}
Ferrara, E., Varol, O., Davis, C., Menczer, F., and Flammini, A. (2016b).
\newblock The rise of social bots.
\newblock {\em Comm. of the ACM\/}, {\bf 59}(7), 96--104.

\bibitem[Fourney {\em et~al.}(2017)Fourney, Racz, Ranade, Mobius, and
  Horvitz]{fourney2017geographic}
Fourney, A., Racz, M.~Z., Ranade, G., Mobius, M., and Horvitz, E. (2017).
\newblock Geographic and temporal trends in fake news consumption during the
  2016 us presidential election.
\newblock In {\em CIKM\/}, volume~17.

\bibitem[Freitas {\em et~al.}(2015)Freitas, Benevenuto, Ghosh, and
  Veloso]{freitas2015reverse}
Freitas, C., Benevenuto, Ghosh, and Veloso, A. (2015).
\newblock Reverse engineering socialbot infiltration strategies in twitter.
\newblock In {\em ASONAM\/}.

\bibitem[Gibson and McAllister(2006)Gibson and McAllister]{gibson2006does}
Gibson, R.~K. and McAllister, I. (2006).
\newblock Does cyber-campaigning win votes? online communication in the 2004
  australian election.
\newblock {\em Journal of Elections, Public Opinion and Parties\/}, {\bf
  16}(3), 243--263.

\bibitem[Gonz{\'a}lez-Bail{\'o}n {\em et~al.}(2011)Gonz{\'a}lez-Bail{\'o}n,
  Borge-Holthoefer, Rivero, and Moreno]{gonzalez2011dynamics}
Gonz{\'a}lez-Bail{\'o}n, S., Borge-Holthoefer, J., Rivero, A., and Moreno, Y.
  (2011).
\newblock The dynamics of protest recruitment through an online network.
\newblock {\em Scientific reports\/}, {\bf 1}, 197.

\bibitem[Gonz{\'a}lez-Bail{\'o}n {\em et~al.}(2013)Gonz{\'a}lez-Bail{\'o}n,
  Borge-Holthoefer, and Moreno]{gonzalez2013broadcasters}
Gonz{\'a}lez-Bail{\'o}n, S., Borge-Holthoefer, J., and Moreno, Y. (2013).
\newblock Broadcasters and hidden influentials in online protest diffusion.
\newblock {\em American Behavioral Scientist\/}, {\bf 57}(7), 943--965.

\bibitem[Howard(2006)Howard]{howard2006new}
Howard, P. (2006).
\newblock {\em New media campaigns and the managed citizen\/}.

\bibitem[Hwang {\em et~al.}(2012)Hwang, Pearce, and Nanis]{hwang2012socialbots}
Hwang, T., Pearce, I., and Nanis, M. (2012).
\newblock Socialbots: Voices from the fronts.
\newblock {\em Interactions\/}, {\bf 19}(2), 38--45.

\bibitem[Kloumann {\em et~al.}(2012)Kloumann, Danforth, Harris, Bliss, and
  Dodds]{kloumann2012positivity}
Kloumann, I.~M., Danforth, C.~M., Harris, K.~D., Bliss, C.~A., and Dodds, P.~S.
  (2012).
\newblock Positivity of the english language.
\newblock {\em PloS one\/}, {\bf 7}(1), e29484.

\bibitem[Kollanyi {\em et~al.}(2016)Kollanyi, Howard, and
  Woolley]{kollanyi2016bots}
Kollanyi, B., Howard, P.~N., and Woolley, S.~C. (2016).
\newblock Bots and automation over twitter during the first us presidential
  debate.

\bibitem[Kudugunta and Ferrara(2018)Kudugunta and Ferrara]{kudugunta2018deep}
Kudugunta, S. and Ferrara, E. (2018).
\newblock Deep neural networks for bot detection.
\newblock {\em Information Sciences\/}, {\bf 467}(October), 312--322.

\bibitem[Loader and Mercea(2011)Loader and Mercea]{loader2011networking}
Loader, B.~D. and Mercea, D. (2011).
\newblock Networking democracy? social media innovations and participatory
  politics.
\newblock {\em Inf. Commun. Soc\/}, {\bf 14}(6).

\bibitem[Messias {\em et~al.}(2013)Messias, Schmidt, Oliveira, and
  Benevenuto]{messias2013you}
Messias, J., Schmidt, L., Oliveira, R., and Benevenuto, F. (2013).
\newblock You followed my bot! transforming robots into influential users in
  twitter.
\newblock {\em First Monday\/}, {\bf 18}(7).

\bibitem[Metaxas and Mustafaraj(2012)Metaxas and Mustafaraj]{metaxas2012social}
Metaxas, P.~T. and Mustafaraj, E. (2012).
\newblock Social media and the elections.
\newblock {\em Science\/}, {\bf 338}(6106).

\bibitem[Monsted {\em et~al.}(2017)Monsted, Sapiezynski, Ferrara, and
  Lehmann]{monsted2017evidence}
Monsted, B., Sapiezynski, P., Ferrara, E., and Lehmann, S. (2017).
\newblock Evidence of complex contagion of information in social media: An
  experiment using twitter bots.
\newblock {\em PLOS ONE\/}, {\bf 12}(9), 1--12.

\bibitem[Pennycook and Rand(????)Pennycook and Rand]{Pennycook2017b}
Pennycook, G. and Rand, D.~G. (????).
\newblock {Assessing the Effect of ``Disputed'' Warnings and Source Salience on
  Perceptions of Fake News Accuracy}.

\bibitem[Raghavan {\em et~al.}(2007)Raghavan, Albert, and Kumara]{raghavan2007}
Raghavan, U.~N., Albert, R., and Kumara, S. (2007).
\newblock Near linear time algorithm to detect community structures in
  large-scale networks.
\newblock {\em Physical review E\/}, {\bf 76}(3), 036106.

\bibitem[Ratkiewicz {\em et~al.}(2011a)Ratkiewicz, Conover, Meiss,
  Gon{\c{c}}alves, Flammini, and Menczer]{ratkiewicz2011detecting}
Ratkiewicz, J., Conover, M., Meiss, M.~R., Gon{\c{c}}alves, B., Flammini, A.,
  and Menczer, F. (2011a).
\newblock Detecting and tracking political abuse in social media.
\newblock {\em ICWSM\/}, {\bf 11}, 297--304.

\bibitem[Ratkiewicz {\em et~al.}(2011b)Ratkiewicz, Conover, Meiss,
  Gon{\c{c}}alves, Patil, Flammini, and Menczer]{ratkiewicz2011truthy}
Ratkiewicz, J., Conover, M., Meiss, M., Gon{\c{c}}alves, B., Patil, S.,
  Flammini, A., and Menczer, F. (2011b).
\newblock Truthy: mapping the spread of astroturf in microblog streams.
\newblock In {\em 20th WWW Conf.}, pages 249--252.

\bibitem[Savage {\em et~al.}(2016)Savage, Monroy-Hernandez, and
  H{\"o}llerer]{savage2016botivist}
Savage, S., Monroy-Hernandez, A., and H{\"o}llerer, T. (2016).
\newblock Botivist: Calling volunteers to action using online bots.
\newblock In {\em 19th CSCW\/}.

\bibitem[Shirky(2011)Shirky]{shirky2011political}
Shirky, C. (2011).
\newblock The political power of social media: Technology, the public sphere,
  and political change.
\newblock {\em Foreign affairs\/}, pages 28--41.

\bibitem[Shorey and Howard(2016)Shorey and Howard]{shorey2016automation}
Shorey, S. and Howard, P.~N. (2016).
\newblock Automation, algorithms, and politics: A research review.
\newblock {\em Int. J Comm.}, {\bf 10}.

\bibitem[Stella {\em et~al.}(2018)Stella, Ferrara, and
  De~Domenico]{stella2018bots}
Stella, M., Ferrara, E., and De~Domenico, M. (2018).
\newblock Bots increase exposure to negative and inflammatory content in online
  social systems.
\newblock {\em Proceedings of the National Academy of Sciences\/}, page
  201803470.

\bibitem[Subrahmanian {\em et~al.}(2016)Subrahmanian, Azaria, Durst, Kagan,
  Galstyan, Lerman, Zhu, Ferrara, Flammini, and Menczer]{subrahmanian2016darpa}
Subrahmanian, V., Azaria, A., Durst, S., Kagan, V., Galstyan, A., Lerman, K.,
  Zhu, L., Ferrara, E., Flammini, A., and Menczer, F. (2016).
\newblock The darpa twitter bot challenge.
\newblock {\em Computer\/}, {\bf 49}(6).

\bibitem[Tufekci(2014)Tufekci]{tufekci2014big}
Tufekci, Z. (2014).
\newblock Big questions for social media big data: Representativeness, validity
  and other methodological pitfalls.
\newblock {\em ICWSM\/}.

\bibitem[Tufekci and Wilson(2012)Tufekci and Wilson]{tufekci2012social}
Tufekci, Z. and Wilson, C. (2012).
\newblock Social media and the decision to participate in political protest:
  Observations from tahrir square.
\newblock {\em Journal of Communication\/}, {\bf 62}(2), 363--379.

\bibitem[Varol {\em et~al.}(2014)Varol, Ferrara, Ogan, Menczer, and
  Flammini]{varol2014evolution}
Varol, O., Ferrara, E., Ogan, C.~L., Menczer, F., and Flammini, A. (2014).
\newblock Evolution of online user behavior during a social upheaval.
\newblock In {\em Proceedings of the 2014 ACM conference on Web science\/}.

\bibitem[Varol {\em et~al.}(2017a)Varol, Ferrara, Menczer, and
  Flammini]{varol2017early}
Varol, O., Ferrara, E., Menczer, F., and Flammini, A. (2017a).
\newblock Early detection of promoted campaigns on social media.
\newblock {\em EPJ Data Science\/}, {\bf 6}(13).

\bibitem[Varol {\em et~al.}(2017b)Varol, Ferrara, Davis, Menczer, and
  Flammini]{varol2016online}
Varol, O., Ferrara, E., Davis, C., Menczer, F., and Flammini, A. (2017b).
\newblock Online human-bot interactions: Detection, estimation, and
  characterization.
\newblock In {\em ICWSM\/}, pages 280--289.

\bibitem[Warriner {\em et~al.}(2013)Warriner, Kuperman, and
  Brysbaert]{warriner2013norms}
Warriner, A.~B., Kuperman, V., and Brysbaert, M. (2013).
\newblock Norms of valence, arousal, and dominance for 13,915 english lemmas.
\newblock {\em Behavior research methods\/}, {\bf 45}(4), 1191--1207.

\bibitem[Wilson {\em et~al.}(2005)Wilson, Wiebe, and
  Hoffmann]{wilson2005recognizing}
Wilson, T., Wiebe, J., and Hoffmann, P. (2005).
\newblock Recognizing contextual polarity in phrase-level sentiment analysis.
\newblock In {\em Proceedings of the conference on human language technology
  and empirical methods in natural language processing\/}, pages 347--354.
  Association for Computational Linguistics.

\bibitem[Woolley and Howard(2016)Woolley and Howard]{woolley2016automation}
Woolley, S.~C. and Howard, P.~N. (2016).
\newblock Automation, algorithms, and politics: Introduction.
\newblock {\em Int. Journal of Commun.}, {\bf 10}.

\end{thebibliography}
\end{document}